\documentstyle[11pt]{article}
\begin{document}
\hoffset = -1truecm
\voffset = -2truecm
\title{\bf Higher dimensional Lax pairs \\ of lower dimensional chaos and turbulence systems}
\author{Sen-yue Lou\thanks{Email: sylou@mail.sjtu.edu.cn}\\
\it \footnotesize \it CCAST (World Laboratory), PO Box
8730, Beijing 100080, P. R. China\\
\it \footnotesize \it Physics Department
of Shanghai Jiao
Tong University,
Shanghai 200030, P. R. China\\
\footnotesize \it Physics Department of Ningbo University, Ningbo
315211, P. R. China}
\date{}

\maketitle

\begin{abstract}
In this letter, a definition of the higher dimensional Lax pair
for a lower dimensional system which may be a chaotic system is
given. A special concrete (2+1)-dimensional Lax pair for a general
(1+1)-dimensional three order autonomous partial differential
equation is studied. The result shows that any (1+1)-dimensional
three order semi-linear autonomous system (no matter it is
integrable or not) possesses infinitely many (2+1)-dimensional Lax
pairs. Especially, every solution of the KdV equation and the
Harry-Dym equation with their space variable being replaced by the
field variable can be used to obtain a (2+1)-dimensional Lax pair
of any three order (1+1)-dimensional semi-linear equation.
\end{abstract}
\vskip.1in

In the nonlinear mathematical physics, if an (n+1)-dimensional
nonlinear system can be considered as a consistent condition of an
(n+1)-dimensional linear system, then many types of interesting
properties of the nonlinear system can be obtained by analyzing
the linear system. The linear system is called as the Lax pair of
the original nonlinear system$\cite{AC}$.

In this letter, we use the concept of the Lax pair under a more
general meaning. If the compatible condition
\begin{eqnarray}
[L_1,\ L_2]\psi \equiv (L_1L_2-L_2L_1)\psi=0,\
\end{eqnarray}
of a pair of linear equations
\begin{eqnarray}
&& L_1(x_1,x_2,...,x_n, t, u(x_1,x_2,...,x_m, t))\psi (x_1,x_2,...,x_n, t)\equiv L_1\psi =0, \\
&& L_2(x_1,x_2,...,x_n, t, u(x_1,x_2,...,x_m, t))\psi (x_1,x_2,...,x_n, t)\equiv L_2\psi =0,\qquad m<n,
\end{eqnarray}
is only a $\psi$-independent nonlinear equation of
$u(x_1,x_2,...,x_m, t)\equiv u$
\begin{eqnarray}
F(u)=0,
\end{eqnarray}
then we call the system (2) and (3) the (n+1)-dimensional Lax pair
of the (m+1)-dimensional nonlinear equation (4) (or equivalently
(1)).

To find some nontrivial examples, we restrict ourselves to
discussion the case for
\begin{eqnarray}
&&L_1\equiv \frac{\partial^2}{\partial x\partial y}-F_{y}(t,\ y,\ u, u_x),\qquad u=u(x,t)\\
&&L_2\equiv \frac{\partial}{\partial t} +
\frac{\partial^3}{\partial x^3}-G(t,\ y,\ u,\ u_x,\
u_{xx})\frac{\partial}{\partial x},
\end{eqnarray}
where $F\equiv F(t,\ y,\ u, u_x),\ G\equiv G(t,\ y,\ u,\ u_x,\
,u_{xx})$ should be selected appropriately such that Eq. (1) is
only a (1+1)-dimensional partial differential equation of $u$. For
the notation simplicity later, the undetermined function  in the
equation (5) is written as a partial derivative form.

In the next step we try to find some autonomous evolution models
\begin{eqnarray}
u_t=H_0(u,\ u_x,\ u_{xx}, \ u_{xxx},\ u_{xxxx},\ ...,)
\end{eqnarray}
 which are given by the compatibility conditions of (2) and (3) with (5) and (6)
\begin{eqnarray}
\psi_{xyt}=\psi_{txy}
\end{eqnarray}
or equivalently (1) with (5) and (6).

Using the relations (5), (6) and (7), the compatibility condition
(8) has the following form
\begin{eqnarray}
W_1(t,y,u,u_x,...)\psi+W_2(t,y,u,u_x,...)\psi_x+(3F_{u_x}u_{xx}-G+3F_uu_x)_y\psi_{xx}=0,
\end{eqnarray}
where $W_1(t,y,u,u_x,...)\equiv W_1$ and $W_2(t,y,u,u_x,...)\equiv
W_2$ are two complicated functions of the indicated variables and
$F_{u_x}$ and $F_u$ denotes the partial derivatives of $F$ with
respect to $u_x$ and $u$. The $\psi$-independent condition of (8)
requires that three terms of (9) should all be zero. Vanishing the
last term of (9) yields
\begin{eqnarray}
G=3F_{u_x}u_{xx}+3F_uu_x+G_0(t,u,u_x,u_{xx}),
\end{eqnarray}
where $G_0(t,u,u_x,u_{xx})$ is a $y$-independent arbitrary
function of the indicated variables. Substituting (10) into (9),
we find that $W_2$ becomes zero identically while $W_1=0$ is
simplified to
\begin{eqnarray}
W_{11}(t,y,u,u_x,...,u_{k,x})+F_{yu_x}H_{0u_{k,x}}u_{k+1,x}=0,\qquad
k>3,
\end{eqnarray}
where $u_{k,x}$ denotes the $k$th order derivative of $u$ with
respect to $x$, $k$ is the highest derivative order of $u$
included in $H_0(u,\ u_x,\ u_{xx}, \ u_{xxx},\ u_{xxxx},\
...,)\equiv H_0$ and $W_{11}$ is $u_{K,x} (K\geq k+1)$ independent
complicated function. Because (11) should be true for the general
solution $u$ of (7), two terms of (11) should all be zero. For
$F_{yu_x}\neq0$, vanishing second term of (11), we have the result
that $H$ is $u_{k,x}$ independent for $k\geq 4$.

For $k=3$, (8) with (10) has the form
\begin{eqnarray}
W_{11}(t,y,u,u_x,u_{xx},u_{xxx})+F_{yu_x}(H_{0u_{xxx}}+1)u_{xxxx}=0.
\end{eqnarray}
Vanishing the second term of (12) yields the result that the only
possible form of $H_0$ can be written as
\begin{eqnarray}
H_0=-u_{xxx}+H(u,\ u_x,\ u_{xx})
\end{eqnarray}
for $F_{yu_x}\neq0$. Substituting (13) back to (12) yields
\begin{eqnarray}
W_{12}(t,y,u,u_x,u_{xx})+(F_{yu_x}H_{u_{xx}}-3F_yF_{u_x}+3F_{yuu_x}u_x+3F_{yu_xu_x}u_{xx}-F_yG_{0u_{xx}})u_{xxx}=0.
\end{eqnarray}
Because $W_{12}(t,y,u,u_x,u_{xx})$ is $u_{xxx}$ independent, the
second term of (14) should be also zero. Vanishing the
coefficients of $u_{xxx}$ of (14) and partial integrating once
with respect to $u_{xx}$ we obtain
\begin{eqnarray}
\frac32F_{yu_xu_x}u_{xx}^2+3(F_{yuu_x}u_x-F_yF_{u_x})u_{xx}-F_yG_0+F_{yu_x}H+F_1(t,y,u,u_x)=0,
\end{eqnarray}
where $F_1(t,y,u,u_x)\equiv F_1$ is an integrating function.
Because $F$ and $F_1$ in (15) are $y$-dependent while all other
functions in (15) are all $y$-independent, we can conclude that
the only possible solution of (15) has the form
\begin{eqnarray}
G_0=\frac32\Gamma_2(t,u,u_x)u_{xx}^2+\Gamma_3(t,u,u_x)u_{xx}+\Gamma_1(t,u,u_x)H+\Gamma_4(t,u,u_x),
\end{eqnarray}
for general $H$ with
\begin{eqnarray}
&& F_{yu_x}=F_y\Gamma_1(t,u,u_x),\\
&&F_{yu_xu_x}=F_y\Gamma_2(t,u,u_x),\\
&&3(F_{yuu_x}u_x-F_yF_{u_x})=F_y\Gamma_3(t,u,u_x),\\
&& F_1(t,y,u,u_x)=F_y\Gamma_4(t,u,u_x),
\end{eqnarray}
and $\Gamma_1(t,u,u_x),\ \Gamma_2(t,u,u_x),\ \Gamma_3(t,u,u_x)$
and $ \Gamma_4(t,u,u_x)$ being four unknown functions which will
be determined later.

According to the compatibility condition $F_{uu_x}=F_{u_xu}$ and
the fact that $F$ is the only $y$-dependent function, we can
obtain the unique possible solution of (17)-(19):
\begin{eqnarray}
&&\Gamma_1(t,u,u_x)=u_x^{-1},\\
&&\Gamma_2(t,u,u_x)=0,\\
&&\Gamma_3(t,u,u_x)=-3G_{1u_x}(t,u,u_x),\\
&&F=G_1(t,u,u_x)+u_xF_0(t,y,u),\\
 &&F_{0u}(t,y,u)=\frac12F_0(t,y,u)^2+\Gamma_5(t,u),
\end{eqnarray}
where $G_1(t,u,u_x)$ is an arbitrary function of $\{t,u,u_x\}$ and
$\Gamma_5(t,u)$ is an arbitrary function of $\{t,u\}$. Using the
relations (16), (21)-(25), Eq. (12) is simplified to
\begin{eqnarray}
&&\frac{F_{0ty}(t,y,u)}{F_{0y}(t,y,u)}-[2u_x^3\Gamma_5(t,u)
+\Gamma_4(t,u,u_x)u_x+3G_{1u}(t,u,u_x)u_x^2]F_0(t,y,u)\nonumber\\
&&-u_{xx}[3u_x(2\Gamma_5(t,u)+G_{1uu_x}(t,u,u_x))+\Gamma_{4u_x}(t,u,u_x)
+6G_{1u}(t,u,u_x)+u_x^{-1}\Gamma_4(t,u,u_x)]\nonumber\\
&&-u_x\Gamma_{4u}(t,u,u_x)-2u_x^3\Gamma_{5u}-3u_x^2G_{1uu}(t,u,u_x)=0.
\end{eqnarray}
Because $F_0(t,y,u)$ is $y$-dependent, and other functions in (26)
are $y$ and $u_{xx}$ independent, (26) is true only for
\begin{eqnarray}
&&F_{0ty}(t,y,u)=F_{0y}(t,y,u)(\Gamma_6(t,u)F_0(t,y,u)+\Gamma_7(t,u)),\\
&&3u_x(2\Gamma_5(t,u)+G_{1uu_x}(t,u,u_x))+\Gamma_{4u_x}(t,u,u_x)
+6G_{1u}(t,u,u_x)+u_x^{-1}\Gamma_4(t,u,u_x)=0,\\
&&\Gamma_7(t,u)-u_x\Gamma_{4u}(t,u,u_x)-2u_x^3\Gamma_{5u}-3u_x^2G_{1uu}(t,u,u_x)=0,\\
&&2u_x^3\Gamma_5(t,u)+\Gamma_4(t,u,u_x)u_x+3G_{1u}(t,u,u_x)u_x^2-\Gamma_6(t,u)=0,
\end{eqnarray}
where $\Gamma_6(t,u)$ and $\Gamma_7(t,u)$ are arbitrary functions
of $\{t,u\}$. The general solution of (28)-(30) reads
\begin{eqnarray}
&&\Gamma_4(t,u,u_x)=-3u_xG_{1u}(t,u,u_x)-2u_x^2\Gamma_5(t,u)+u_x^{-1}\Gamma_6(t,u),\\
&&\Gamma_7(t,u)=\Gamma_{6u}(t,u).
\end{eqnarray}
Integrating (27) once with respect to $y$ leads to
\begin{eqnarray}
F_{0t}(t,y,u)=\frac12\Gamma_6(t,u)F_0(t,y,u)^2+\Gamma_{6u}(t,u)F_0(t,y,u)+\Gamma_8(t,u),
\end{eqnarray}
where $\Gamma_8(t,u)$ is a further integrating function. Because
one function, $F_0(t,y,u)$, should satisfy two equations (25) and
(33), the compatibility condition $F_{tu}=F_{ut}$ yields the
constraints
\begin{eqnarray}
\Gamma_8(t,u)=\Gamma_6(t,u)\Gamma_5(t,u)+\Gamma_{6uu}(t,u),
\end{eqnarray}
and
\begin{eqnarray}
\Gamma_{5t}(t,u)=\Gamma_{6uuu}(t,u)+\Gamma_6(t,u)\Gamma_{5u}(t,u)+2\Gamma_5(t,u)\Gamma_{6u}(t,u).
\end{eqnarray}

Finally, we write down the obtained results here. A
(1+1)-dimensional three order equation
\begin{eqnarray}
u_t=-u_{xxx}+H(u,u_x,u_{xx})
\end{eqnarray}
with an arbitrary nonlinear interaction term $H(u,u_x,u_{xx})$
possesses a (2+1)-dimensional Lax pair
\begin{eqnarray}
&&\psi_{xy}-u_xF_{0y}(t,y,u)\psi=0,\\
&&\psi_t+\psi_{xxx}-[3F_0(t,y,u)u_{xx}+3u_x^2F_{0u}(t,y,u)+H_1(t,u,u_x,u_{xx})]\psi_x=0,
\end{eqnarray}
where
\begin{eqnarray}
H_1(t,u,u_x,u_{xx})=u_x^{-1}(H(u,u_x,u_{xx})+\Gamma_6(t,u))-2u_x^2\Gamma_5(t,u).
\end{eqnarray}

In the Lax pair equations (37) and (38), the function
$F_0(t,y,u)\equiv F_0$ is determined by a pair of the consistent
Riccati equations
\begin{eqnarray}
&&F_{0t}(t,y,u)=\frac12\Gamma_6(t,u)F_0(t,y,u)^2
+\Gamma_{6u}(t,u)F_0(t,y,u)+\Gamma_6(t,u)\Gamma_5(t,u)+\Gamma_{6uu},\\
&&F_{0u}(t,y,u)=\frac12F_0(t,y,u)^2 +\Gamma_5(t,u)
\end{eqnarray}
while the consistent condition of (40) and (41) gives the
constraint equation (35) for the functions $\Gamma_6(t,u)$ and
$\Gamma_5(t,u)$.

It is known that for the usual Lax pair of a given nonlinear
model, there may be infinitely many Lax pairs$\cite{LouHu}$. The
similar situation appears for the extended higher dimensional Lax
pairs. For instance, for the given evolution system (36), the
arbitrariness in the selections of the functions $\Gamma_5(t,u)$
and $\Gamma_6(t,u)$ and the solutions of $F_0(t,y,u)$ means that
the evolution equation (36) possesses infinitely many different
kinds of Lax pairs. To give out some more concrete results, we may
fix the functions $\Gamma_6(t,u)$ and $\Gamma_5(t,u)$. The first
interesting selection is
\begin{eqnarray}
\Gamma_6(t,u)=\Gamma_5(t,u).
\end{eqnarray}
Under the selection (42), (36) becomes the well known Korteweg
de-Vries (KdV) equation
\begin{eqnarray}
\Gamma_{5t}(t,u)=\Gamma_{5uuu}(t,u)+3\Gamma_5(t,u)\Gamma_{5u}(t,u).
\end{eqnarray}
with the independent variables $\{t,u\}$ and the linearized system
of (40) and (41) by using the Cole-Hopf transformation is just the
Lax pair of the KdV equation (after neglecting the variable $y$ in
(40) and (41)). Now substituting every known special solution of
the KdV equation and the related solution of the psudopotential
$F_0$ of (40) and (41) into (37) and (38), we get a concrete Lax
pair of the system (36). The simplest trivial solution of the KdV
equation is
\begin{eqnarray}
\Gamma_5(t,u)=\Gamma_6(t,u)=0,\
\end{eqnarray}
and the related solution of (40) and (41) reads
\begin{eqnarray}
F_0(t,y,u)=-\frac2{u+q(y)},
\end{eqnarray}
where $q(y)$ is an arbitrary function of $y$.

The second interesting selection is
\begin{eqnarray}
\Gamma_6(t,u)=\Gamma_5^{-1/2}(t,u).
\end{eqnarray}
In this case, (36) becomes the well known Harry-Dym (HD) equation
\begin{eqnarray}
\Gamma_{5t}(t,u)=[\Gamma_{5}^{-1/2}(t,u)]_{uuu}.
\end{eqnarray}
with the independent variables $\{t,u\}$ and the linearized system
of (40) and (41) is the Lax pair of the HD equation. That means
substituting every known special solution of the HD equation and
the related solution of the psudopotential $F_0$ of (40) and (41)
into (37) and (38) will yield a concrete Lax pair of the
semi-linear system (36).

Generally, the evolution equation (36) is nonintegrable under the
traditional meanings. In Ref. $\cite{SSY}$, the authors had
claimed that there exist only six nonequivalent three order
semi-linear integrable models under the usual meanings. In other
words, the evolution equation (36) is chaotic or turbulent for
most of the selections of $H$. The results of this letter show us
that no matter of the model (36) is integrable or not, it may have
infinitely many (2+1)-dimensional Lax pairs. A simple special
nonintegrable example of (36) is the so-called KdV-Burgers (KdVB)
equation$\cite{KdVB}$,
\begin{eqnarray}
u_{t}+uu_{x}-\nu u_{xx}+ u_{xxx}=0.
\end{eqnarray}
The KdVB is one of the possible candidate to describe the
turbulence phenomena in fluid physics and plasma
physics$\cite{turbulence, KdVB}$. Though one has not yet find any
(1+1)-dimensional Lax pair of the turbulence system KdVB equation,
we can do find many types of (2+1)-dimensional Lax pairs.
Especially, every solution of the KdV equation and the HD equation
can be used to form a (2+1)-dimensional Lax pair of the KdVB
equation.

In summary, a lower dimensional nonlinear system may have some
(perhaps infinitely many) types of higher dimensional Lax pairs.
In terms of a special example we show that any three order
semi-linear equation (no mater it is integrable or not) possesses
infinitely many Lax pairs. The next interesting important problem
should be studied in the future work may be: What kinds of
information about lower dimensional systems especially the lower
dimensional turbulent and chaotic systems can be obtained from
some types of special higher dimensional Lax pairs?

\vskip.2in The work was supported by the Outstanding Youth
Foundation and the National Natural Science Foundation of China
(Grant. No. 19925522), the Research Fund for the Doctoral Program
of Higher Education of China (Grant. No. 2000024832) and the
Natural Science Foundation of Zhejiang Province, China. The author
is in debt to thanks the helpful discussions with the professors
Q. P. Liu, G-x Huang and C-p Sun and the Drs. X-y Tang, S-l Zhang,
C-l Chen and B. Wu.

\end{document}